\documentclass[10pt]{article}
\newtheorem{lemma}{Lemma}
\newtheorem{theorem}{Theorem}
\newtheorem{definition}{Definition}

\title{Towards Applicative Relational Programming}
\author{H. Ibrahim and M.H. van Emden}
\date{16 March 1992}

\begin{document}
\maketitle

\begin{abstract}
Functional programming comes in two flavours: one where ``functions are
first-class citizens'' (we call this {\em applicative}) and one which is
based on equations (we call this {\em declarative}). 
In relational programming clauses play the role of equations. Hence
Prolog is declarative. 
The purpose of this paper is to provide in relational programming 
a mathematical basis for the relational analog of applicative functional 
programming.
We use the cylindric semantics of first-order logic due to Tarski and
provide a new notation for the required cylinders that we call {\em tables}. 
We define the {\em Table/Relation Algebra} with operators sufficient to 
translate Horn clauses into algebraic form.
We establish basic mathematical properties of these operators.
We show how relations can be first-class citizens, 
and devise mechanisms for modularity, for local scoping of predicates, 
and for exporting/importing relations between programs.
\end{abstract}

\section{Applicative versus declarative definitions}

Some functional programming systems are applicative; others declarative.
Relational programming, on the other hand, only exists in declarative form.
In this section we explain how we use this terminology,
and argue that relational
programming should also have both declarative and applicative forms.

\subsection{In functional programming}

There are two ways for defining functions in functional programming,
one using $\lambda$-expressions and the other using equations.
Consider for example the higher-order function ``{\em twice}''.
It can be denoted by the $\lambda$-expression 
``$\lambda f.\lambda x.f(f x)$'';
we call this an applicative definition.
This function can also be
defined in a declarative style, that is, by asserting as true
certain equations, as follows:
\begin{verbatim}
twice F = g(F)
g(F) X = F (F X)
\end{verbatim}

The characteristics of each style can be summarized in the first two 
columns of table \ref{fun-rel-chars}.
In the applicative style, functions can be results of functions, can be
bound to variables, and so on. Hence they are much like other types of
values. This is not the case in the declarative style. As a result, it is
sometimes said that in the applicative style, functions are
``first-class citizens.''
\begin{table}
\centerline{
\begin{tabular}{c||c|c}
$Characteristics - Paradigms$ & $functional$ & $relational$     \\ \hline\hline
            $implicit$ &  &  \\ 
            {\em element at a time} & $equations$ & {\em definite clauses} \\ 
            $declarative$ &  &  \\ \hline
            $explicit$ &  &   \\ 
            {\em whole function/relation} & $\lambda-calculus$ & ?  \\ 
            $applicative$ &  &   \\ \hline
\end{tabular}}
\caption{Classifying functional and relational definitions.}
\label{fun-rel-chars}
\end{table}
Furthermore, it can be shown that the two styles have complementary strengths,
namely local scoping and modularity in the applicative one, and a natural
expression of recursion and selection in the declarative one.
Thus it is valuable for a programmer to have both
available and to be able to switch effortlessly between the two.
In functional programming,
the theory of such a combination has been developed and practical
applications have been reported \cite{chn87,rch90}

\subsection{In relational programming}

``Relational programming'' is a natural counterpart for functional programming,
with relations instead of function as basic entities.
Logic programming, the best developed form of relational programming,
is exclusively based on Horn clauses, which constitute a declarative paradigm.
As the third column of table \ref{fun-rel-chars} suggests, a major problem in
relational programming is the absence of an applicative language in which
relations can be first-class citizens.
This paper develops the basis for such a language, keeping in mind the 
importance of it being first-order, and easily intertranslatable.  

\section{The meaning of Horn clauses}

As basis for applicative relational programming, we use an algebraic
view of the meaning of Horn clauses. The following example serves as
introduction.
Consider the clause for relational composition:
$p(X,Z) \leftarrow q(X,Y), r(Y,Z). \label{pqr}$
Relation $p$ is defined by means of operations on $q$, $r$, and the tuples
of variables. We make these operations explicit by appealing to cylindric
set algebra \cite{hmtan81,hmt85}.
Accordingly,
a Horn clause can be interpreted as a relational inclusion 
where the right-hand side is a projection of the intersection of cylinders on 
the relations in the condition part.
Consider, for example,
the geometric interpretation of the clause as illustrated in 
figure \ref{cylinders}.

\begin{figure}[htb]
\caption{A geometric interpretation of relational composition: $p$
contains the projection of the intersection of the cylinders on $q$ and
$r$.}
\label{cylinders}
\end{figure}

This can be expressed by a formula using the operators of $TRA$ introduced
in this paper. As a preview, we list the formula here:
$p \supset (X,Z) / (q:(X,Y) \cap r:(Y,Z)),$
where ``:'' is relational application yielding a cylinder, 
``$\cap$'' is intersection, and ``/'' is relational projection.
To allows local scoping through $\lambda$-binding,
we want to have $p$, $q$, and $r$ as relational variables.
This also allows minimization operators, such as $\mu$ \cite{ddr73}, 
to apply as well.
The result will be relation-valued expressions in a first-order setting.
  
\subsection{Relational semantics for first-order logic.}

{\em Truth-functional semantics} for first-order logic is an assignment of 
a truth value to a {\em closed} formula, where this
assignment is relative to a given interpretation that assigns meanings to
constants, function symbols, and predicate symbols.
{\em Relational semantics} assigns a meaning to a formula that may have free
variables; i.e. an {\em open} formula. It gives as meaning
the relation consisting of the tuples of individuals that, if
assigned to the free variables in the formula, would give
a true closed formula according to truth-functional semantics.

For example, the meaning with respect to an interpretation $I$
of the formula $q(X,Y)$ under relational semantics is the binary
relation
$$r=\{(\alpha ,\beta )|q(\alpha , \beta ) \>
  \hbox{is variable-free and true in} \> I\}$$
For the benefit of further examples, suppose that the universe of
discourse is $\{a,b,c\}$\footnote{The above footnote applies here also.}
and that $r=\{(a,b),(b,c),(c,a)\}$.

Note that the meaning of a relation under relational semantics
is invariant under renaming of free variables.
As a result, relational semantics is not a homeomorphism between the
algebra of formulas and the algebra of relations. For example,
if it were a homeomorphism, then the meaning of
$q(X,Y) \wedge q(Y,Z)$ would be the intersection of the meanings of
the conjuncts. Hence it would be $r \cap r = r$. However, it should be the
composition of $r$ with itself according to the algebra of binary
relations.

Therefore we need something else: cylindric semantics, introduced in the
following.

\paragraph{Tarski's cylinders.}
Tarski introduced a device that leads to a semantics for formulas that
{\em is} a homeomorphism. Consider relations that are subsets of
$D_1 \times \cdots \times D_n$ (call this the {\em domain product}) and
consider
a relation $b$ (call it the {\em base}) that is a subset of
$D_{j_1} \times \cdots \times D_{j_k}$ where
the {\em selector}, $J=\{j_1, \ldots ,j_k\}$ is a subset of
$\{1,\ldots ,n\}$.

Then
$\pi ^{-1} _J(b) = \{(x_1,\ldots ,x_n)| x_i \in D_i, i=1,\ldots ,n
  \hbox{ and } (x_{j_1},\ldots ,x_{j_k}) \in b\}$
is the {\em cylinder} with base $b$, domain product
$D_1\times \cdots \times D_n$, and selector $J$.
A cylinder, being a set of tuples of the same length, is always a
relation. But a relation is not always a cylinder.

For example, take a domain product with $n$ equal to 3,
$D_1 = D_2 = D_3 = \{a,b,c\}$, with selector equal to $\{1,2\}$, and with
base $r=\{(a,b),(b,c),(c,a)\}$. Then the cylinder specified by these
properties is 
\begin{eqnarray} \label{exCyl1}
\{ \;\; (a,b,a), & (a,b,b), & (a,b,c), \nonumber \\
        (b,c,a), & (b,c,b), & (b,c,c),  \\
        (c,a,a), & (c,a,b), & (c,a,c) \;\; \} \nonumber
\end{eqnarray}

Rather than to assign to a formula as meaning a relation according to
relational semantics, Tarski assigns a {\em cylinder} on this relation.
He assumes an enumeration of all variables in the language. This gives a
set $J = \{j_1, \ldots ,j_k\}$ of integers for a formula with $k$ free
variables. The semantics according to Tarski then assigns to the formula
as meaning the cylinder determined by $J$ as selector on
the base that is the relation obtained by
relational semantics. Such a semantics {\em is} a homeomorphism where
conjunction corresponds to intersection. Other syntactic constructs
correspond to other operations in what Tarski calls ``cylindric set
algebra.''

To return to our example, suppose $X$, $Y$, and $Z$ are the only
variables in the language (hence $n=3$), and are
enumerated in this order. Then the free variables of
$q(X,Y)$ correspond to the selector $J = \{1,2\}$. According to Tarski,
the meaning of $q(X,Y)$ is then cylinder \ref{exCyl1}. The free variables
of $q(Y,Z)$ correspond to the selector $J = \{2,3\}$. According to
cylindric semantics, the meaning of $q(Y,Z)$ is then 
\begin{eqnarray} \label{exCyl2}
\{ \;\; (a,a,b), & (b,a,b), & (c,a,b), \nonumber \\
        (a,b,c), & (b,b,c), & (c,b,c), \\
        (a,c,a), & (b,c,a), & (c,c,a) \;\; \} \nonumber
\end{eqnarray}

The meaning of $q(X,Y) \wedge q(Y,Z)$ is the intersection of these two
cylinders, that is, $\{(a,b,c),(b,c,a),(c,a,b)\}$. According to Tarski,
the meaning of $\exists Y. q(X,Y) \wedge q(Y,Z)$ is obtained by projecting
this with selector $J = \{1,3\}$. The result is
$\{(a,c),(b,a),(c,b)\}$, which is the relational composition of $r$ with
itself.

In this paper we define an algebra operating on cylinders and relations
that allows us to translate to algebraic form the definite clauses
of Prolog. We call it {\em TRA}, for Table/Relation Algebra.
The current example serves to give a preview of
this translation. Let a relational composition be defined by the clause
$$p(X,Z) \leftarrow q(X,Y),q(Y,Z).$$
In classical syntax this is
$$(\exists Y. q(X,Y) \wedge q(Y,Z)) \supset p(X,Z).$$
As we just observed, the condition has as meaning the projection on
(X,Z) of the intersection of the cylinders denoted by $q(X,Y)$ and
$q(Y,Z)$. In general, we shall see that a definite clause asserts,
interpreted in terms of TRA, that
{\em the meaning of the conclusion includes the projection (determined
by the variables in the conclusion) of the intersection of the
cylinders that are the meanings of the conditions.}

\subsection{Cylinders defined as tables.}
It is rare to be able to list, as a set of tuples, a cylinder that is a meaning
of a formula; the language usually has infinitely many variables,
so that the tuples are very long. Moreover, there are often infinitely
many tuples in the cylinder. But there is no problem specifying such a
cylinder, as we only need to specify the base relation and the selector
(the domain product is usually implicitly understood).

For example, cylinder \ref{exCyl1} can be specified by the set of
substitutions
$$\{\{X=a,Y=b\},\{X=b,Y=c\},\{X=c,Y=a\}\},$$
as this specifies that the base relation of the cylinder is
$\{(a,b),(b,c),(c,a)\}$ and that the selector is $\{1,2\}$, which
corresponds to the variables $X$ and $Y$ in the enumeration.

Sets of substitutions of this type, where the substituted variables in
each are the same, play a central role in this paper.
Observe, in the first place, that a table is a natural notation for such
a set of substitutions. The right-hand sides make up the lines of the
table, while the left-hand sides need not be repeated, hence can be the
headings of the table's columns; see table \ref{cyl-to-tab}.
Because of this, we call such sets of substitutions ``tables.''
\begin{table}[htb]
\centerline{\begin{tabular}{|c|c|} \hline
          $X$ & $Y$ \\ \hline
          $a$ & $b$ \\
          $b$ & $c$ \\
          $c$ & $a$ \\ \hline
\end{tabular}}
\caption{A table for cylinder \protect\ref{exCyl1}.}
\label{cyl-to-tab}
\end{table}

In the second place, observe that the set of answers to a Prolog query
is a table. For example, suppose that the logic program
$P$ is the set of facts
$\{q(a,b),q(b,c),q(c,a)\}.$ The query $\hbox{?-- } q(X,Y)$ produces 
table \ref{cyl-to-tab} as set of answer substitutions.

\section{The Table/Relation Algebra (TRA)}

TRA is an algebra of operations on relations, tables, queries, and logic 
programs. It facilitates translation to and from definite clause form, 
and also facilitates scoping and modularity. 

We define an $n$-ary {\em relation} over a Herbrand universe
$H$ to be a set of $n$-tuples of elements of $H$.
Prolog's answer substitutions for the set of variables in a query can be
regarded as a set of equations in {\em solved form}\/; i.e. all
left-hand sides are variables that occur only there \cite{mmn82}.
Restricting the solved forms to have a common set of variables as
left-hand sides makes it natural to present a set of answer substitutions
as a table.
Hence we define an $n$-ary {\em table}
to be a set of sets of equations in solved form,
where each set of equations in the table has the same set
of $n$ variables as left-hand sides.

The reason for calling the concept just defined ``table'' is that these
sets of equations can be most economically represented in print as a
table where the left-hand sides are the headings of the columns
(analogous to the attributes of relational data models) and the
right-hand sides are the entries of the table. A difference
with the relational data model
is that in our concept of table the entries can be terms of any
complexity containing variables without any restriction.

\subsection{Tables as queries on logic programs.}
Tables, as defined above, can be obtained, in logic programming,
as follows:

\begin{definition}
Let $P$ be a logic program, let
$Q$ be a query, and let $T$ be an SLD-tree for $P$ and $Q$. Then the expression
$(Q \hbox{ {\tt where} } P)$ has as value the table of the answer substitutions
associated with all the success leaves of $T$. \label{where}
\end{definition}

When in Prolog a query $Q$ fails for a logic program $P$, there are no
answer substitutions and the table $(Q \hbox{ {\tt where} } P)$ is empty.
We use the symbol $\bot$ (bottom) for all of the empty tables $\{ \}.$
We use another special symbol $\top$ (top) for the
table consisting only of the empty answer substitution
$\{ \{ \} \}$, which results from a successful SLD-derivation starting in
a query with no variables.

The following lemmas follow from the definition of ``{\tt where}''.

\begin{lemma} \label{where2correct}
$\theta  \in (Q \hbox{ {\tt where} } P) \hbox{ implies that }
\theta$ is a correct answer substitution for $\{Q\} \cup P.$
\end{lemma}

\begin{lemma}\label{correct2where}
For every correct answer substitution $\theta$ for $\{Q\} \cup P$, $\exists$
$\eta \in (Q \hbox{ {\tt where} } P)$ such that $\theta$ is an
instance (with respect to the Herbrand universe) of $\eta$.
\end{lemma}

\begin{lemma}
The value of $(Q \hbox{ {\tt where} } P)$ does not depend
on the SLD-tree $T$ in definition \ref{where}.
\end{lemma}

\subsection{Intersection of tables}

Tables are a convenient notation for cylinders.
As the intersection of cylinders is important, we need to
define intersection between tables. 
Let $\theta_1$ and $\theta_2$ be answer substitutions to the queries
$G_1$ and $G_2$ respectively, and consider the query $\hbox{?-- }G_1,G_2$.
When $G_1$ and $G_2$ have a common variable, $\theta_1 \cup \theta_2$
is not in solved form and may not be solvable.
However, as Colmerauer \cite{clm82} observed, the solved form of
$\theta_1 \cup \theta_2$, if it exists, is an answer substitution for 
the query $\hbox{?-- }G_1,G_2$.  Hence:
\begin{definition}
The intersection operation of tables $S$ and $T$
is \vspace{-1ex} 
$$S \cap T
\stackrel{\rm def}{=}
  \{ \phi (s \cup t) \mid s \in S, t \in T, and \; \psi(s \cup t)\}.$$
\end{definition}
Here $\psi(s)$ means that its argument set of equations is solvable;
$\phi(s)$ is only defined when $s$ is solvable and then it denotes
the solved form of $s$.
For example, if relation $q$, as defined in a program $P$, 
is $\{(a,b),(b,c),(c,d),(d,e)\}$, then the result of the
query $((\leftarrow q(X,Y)) \hbox{ {\tt where} } P) \cap 
((\leftarrow q(Y,Z)) \hbox{ {\tt where} } P)$ is table \ref{q-tab}.
\begin{table}[htb]
\centerline{\begin{tabular}{|c|c|c|} \hline
          $X$ & $Y$ & $Z$ \\ \hline
          $a$ & $b$ & $c$ \\
          $b$ & $c$ & $d$ \\
          $c$ & $d$ & $e$ \\ \hline
\end{tabular}}
\caption{A table from intersection of two tables.}
\label{q-tab}
\end{table}

The reason for choosing the name ``intersection'' and the symbol ``$\cap$''
is given by the following
\begin{theorem}
For all tables S and T, $S \cap T$ is the set intersection of S and T
regarded as cylinders (hence relations, hence sets).
\end{theorem}
It is easy to verify the following properties about the intersection of tables:
\begin{theorem}
$\cap$ is associative and commutative. $\cap$ has a unique
null element, which is $\bot$ and a unique unit element, which is $\top$.
\end{theorem}

\begin{theorem}
For any program $P$ and goal statements $G_1$, $G_2$ and 
$G$ which consists of all goals in $G_1$ and $G_2$, we have
$$(G_1 \hbox{ {\tt where} } P) \cap (G_2 \hbox{ {\tt where} } P) =
 (G \hbox{ {\tt where} } P).$$
\label{2goals}
\end{theorem} \vspace{-2ex}

\paragraph{Modularity}
The previous theorem suggests using the combination of {\tt where} and $\cap$
as a modular compositional tool, as in the following example:
$$ ((\leftarrow F_1,F_2) \hbox{ {\tt where} } P) \cap 
   ((\leftarrow G) \hbox{ {\tt where} } Q).$$
Thus we see that within the same table it can be specified of each
goal with respect to which program it is defined.

\subsection{From relations to tables}

We often need to get a table out of a relation, rather
than from a program by posing a query.
For this we define the relational {\em application} operator ``:''.
We introduce its definition through a heuristic development.
Let $r$ be the $n$-ary relation $\{(a_1,\ldots,a_n),(b_1,\ldots,b_n),
(c_1,\ldots,c_n)\}$.
A table ``most like'' $r$ can be easily constructed by adding as a heading
a set of $n$ distinct variables, say $X_1,\ldots ,X_n$, as in table 
\ref{tab-like-rel}.
\begin{table}[htb]
\centerline{\begin{tabular}{|c|c|c|} \hline 
          $X_1$ & $\cdots$ & $X_n$ \\ \hline
          $a_1$ & $\cdots$ & $a_n$ \\
          $b_1$ & $\cdots$ & $b_n$ \\ 
          $c_1$ & $\cdots$ & $c_n$ \\ \hline
\end{tabular}}
\caption{A table ``most like'' $r$.}
\label{tab-like-rel}
\end{table}
This table can then be identified using an arbitrary predicate symbol,
say $p$, as follows:
$$(\leftarrow p(X_1,\ldots ,X_n)) \hbox{ {\tt where} }
  \{p(e_1,\ldots ,e_n) \mid (e_1,\ldots ,e_n) \in r \}.$$
The obvious generalization of allowing any terms $t_1, \ldots ,t_n$
instead of the distinct variables suggests
$$r : (t_1,\ldots ,t_n) \stackrel{\rm def}{=}(\leftarrow p(t_1,\ldots ,t_n)) 
\hbox{ {\tt where} } \{p(e_1,\ldots ,e_n) \mid (e_1,\ldots ,e_n) \in r \}.$$
Though correct, its arbitrary and auxiliary predicate $p$ is undesirable.
It is easily verifiable that this definition is equivalent, operationally, to:

\begin{definition}
The application operation of an $n$-ary relation $r$
to a tuple $(t_1,\ldots,t_n)$ of terms is \\
$r : (t_1,\ldots ,t_n) \stackrel{\rm def}{=}
$ \\ \vspace{-2ex}
$$\{\phi (\{t_1=e_1,\ldots ,t_n=e_n\}) \mid
             (e_1,\ldots ,e_n) \in r \hbox{ and }
              \psi (\{t_1=e_1,\ldots ,t_n=e_n\})\}.$$
\end{definition}

\subsection{From tables to relations}

Just as we defined the application operator to get tables from relations,
we define a {\em projection} operation to get relations back from tables.
This operation should not merely discard the table's ``headings''\footnote 
{One reason is, as mentioned before, that our table's contents include
variables, whereas the elements of a relation's tuples are variable-free.},
as suggested in \cite{llmn88}.
Instead, we present another heuristic development to define it properly.

Let $T$ be the previous table \ref{tab-like-rel}.
Now, $\{(X_{1}\theta,\ldots ,X_{n}\theta)\mid \theta \in T\}$ 
is the relation resulting from discarding the heading; i.e. $r$ above.
To generalize, we first let $\{j_1,\ldots ,j_k\}$ be a subset of 
$\{1,\ldots ,n\}$.
Then, $\{(X_{j_1}\theta,\ldots ,X_{j_k}\theta)\mid \theta \in T\}$
is a projection of $r$ over the $(j_1,\ldots ,j_k)$ columns.
Hence, projection is a way of getting a relation from a table.
Secondly, allow any tuple $(t_1,\ldots ,t_k)$ of terms instead of the variables 
$(X_{j_1},\ldots ,X_{j_k})$ and consider $\{(t_{1}\theta,\ldots ,t_{k}\theta)
\mid \theta \in T\}.$ 
Since for an arbitrary table $T$ any $\theta \in T$ may contain variables, 
the result must further be grounded in order to obtain a relation.
Hence the final definition:

\begin{definition}
The projection operation (denoted $/$) of a table $T$ over a tuple 
of terms $(t_1,\ldots,t_n)$ is
\vspace{-1ex}
$$(t_1,\ldots ,t_n)/T \stackrel{\rm def}{=}
  \xi(\{(t_1\theta ,\ldots ,t_n\theta ) \mid \theta \in T\}),$$ \label{project}
where $\xi (x)$ is the set of variable-free instances of the expression $x$.
\end{definition}

\subsection{Are {\em project} and {\em apply} inverses?}

Now that we have operations from tables
to relations and vice versa, one
may wonder whether these are each other's inverses.
The short answer is, in general, ``no'', because
$$((t_1,\ldots ,t_n)/T) : (t_1,\ldots ,t_n)$$
is not always the table $T$. Take, for example, the case that
$t_1,\ldots ,t_n$ have no variables.
Then the above expression is $\top$ whenever $T$ is not $\bot$.
But the absence of variables in $t_1,\ldots,t_n$ is a
rather pathological case. When we add restrictions, we can say
that, in a sense, ``/'' and ``:''
are each other's inverses, as shown by the following theorems.

%
%
\begin{theorem}
For all tables $T$ and all terms $t_1,\ldots ,t_n$ in which all the variables,
and no other ones, in $T$'s heading occur, we have
$$((t_1,\ldots ,t_n)/T) : (t_1,\ldots ,t_n) = \xi(T).$$
\label{exact} \vspace{-5ex}
\end{theorem}

For an inverse in the other direction, compare the $n$-ary relation $r$ with
$$(t_1,\ldots ,t_n)/(r : (t_1,\ldots ,t_n)).$$
That this expression does not always equal $r$ is shown by
$(c,d)/(\{(a,b)\} : (c,d)) = \{\},$
where $a$, $b$, $c$ and $d$ are constants.
This example suggests:

\begin{theorem}
For all $n$-ary relations $r$ and all terms $t_1,\ldots ,t_n$, we have
$$(t_1,\ldots ,t_n)/(r : (t_1,\ldots ,t_n)) \subseteq r.$$
\end{theorem}

However, by strengthening the restrictions, we can have equality instead of
inclusion. 

\begin{theorem}
For all $n$-ary relations $r$ and all distinct variables $x_1,\ldots,x_n$, we
have $$(x_1,\ldots ,x_n)/(r : (x_1,\ldots ,x_n)) = r.$$
\end{theorem}

\subsection{Translation of definite clauses to TRA}

It should be clear now that the operators of TRA correspond closely to the
operations hidden in definite Horn clauses, according to cylindric semantics.
This simplifies the translation of definite clauses to TRA,
and we therefore omit an explicit description of it.

The principle of the translation is that each definite clause states
that {\em the relation denoted by the predicate symbol in the conclusion
includes the projection (on the tuple of the terms in the conclusion) of
the intersection of the tables denoted by the conditions.} The
operations of TRA are general enough to translate the definite clauses
of pure Prolog.

\paragraph{Example.}
We illustrate the translation by an example that is a typical Prolog program.
It gives a quicksort program from an ordinary list to a difference list.

The Prolog program is:
\begin{verbatim}
qsort([ ],U-U).
qsort([X|Xs],U-W) :-
          partition(X,Xs,Y1,Y2),qsort(Y1,U-[X|V]),qsort(Y2,V-W).
\end{verbatim}

The TRA version is:
\begin{eqnarray*}
\lefteqn{qsort \supseteq (([\,], U-U)/\top)} \hspace{1em} \, \\
 & \;\; \cup \; (([X|Xs], U-W)\;\;/ & (((\leftarrow partition(X, Xs, Y1, Y2))
 \hbox{ {\tt where} } P) \\
 &                              & \cap (qsort:(Y1, U-[X|V])) \\
 &                              & \cap (qsort:(Y2, V-W)))).
\end{eqnarray*}

$qsort$ is a relational variable and is not part of the language of clausal
logic. But all terms, and the atomic formula that
is the first argument of {\tt where}, are in clausal logic.

The entire expression states an inclusion between relations and may or may
not be satisfied, depending on the value of {\em qsort}. It may be shown
that there is a least relation as value for {\em qsort} that satisfies
the inclusion. The inclusion serves as definition of this least relation.

The applicative relational program has the following advantages:
\begin{enumerate}
\item
The {\tt where} expression has a table as value and can be replaced by any other
expression with the same value; see item 3. In particular, the 
identifier {\em partition} is local to the {\tt where} expression. The
declarative Horn clause formalism does not provide locality for predicate names.

\item
The identifier {\em qsort} is a variable and is not a predicate symbol.
Hence a minimization operator, like $\mu$ of \cite{ddr73}, can be
applied to the entire inclusion with respect to {\em qsort}. The result
is a relation-valued expression that can, for example, be the operator
argument of the relational application ``:''.
The identifier {\em qsort} is then local to this operator.
In this way any number of levels of locality can be built,
just as in applicative functional programming.

\item
Combining a selection/minimization operator with the
functional $\lambda$-abstraction and application operators
yields an import/export facility for relations between program modules.
As an example, replace $P$ above with
$$ (\lambda order.P) (\nu leq.Orderings),$$ 
where $order$ is a relational variable in $P$ used to
define the order of partitioning, $Orderings$ is a program that
defines different orders, and
``$\nu$'' is the minimization operator that selects one of the relations 
({\em leq})
defined in a program module ({\em Orderings}). The effect of all this is 
exporting the $leq$ relation from program $Orderings$, and importing it into
program $P$ as the order of partitioning, yielding an ascending, or descending,
$qsort$ relation.
\end{enumerate}

\section{Related work}

This paper is a continuation of a line of research
going back as far as the
work of Peirce and of Schroeder in the 19th century on algebra of
relations. These algebras were not adequate to serve as basis for a
semantics for full first-order predicate logic. In the 1930's Tarski
provided cylindric set algebra, which remedied this shortcoming.
The simplicity of definite clauses, although covered by Tarski's work,
suggest the independent treatment given in this paper.

After Tarski's work, the next most important step was the paper by de
Bakker and de Roever \cite{ddr73}. By restricting themselves to binary
relations, they took their starting point before Tarski.
They elucidated the mechanisms of defining binary relations, especially
the use of the minimization operator. 
{\em TRA} facilitates 
extending of the definition techniques of de Bakker and de Roever to include the
type of algebra introduced by Tarski.

There are some relations to recent work that do not seem to be
part of a grand design.
Codd's relational calculus is declarative in spirit, whereas his
relational algebra is applicative in spirit. Both seem to us to be
improved upon by the earlier work by Tarski.
However, TRA can be viewed as a relational algebra counterpart for the 
Datalog query language, when substituted for Codd's relational calculus.
Our {\em where} expressions were inspired by Nait-Abdallah's ions.
The main difference is that Nait-Abdallah's work is syntactic,
concentrating on rewriting rules.

Our work is also related to higher-order logic programming by having
the same objectives, but a different approach; see the
concluding remarks.
Finally, applicative programming constitutes important
related work \cite{sto77}, as mentioned in the introduction.
\vspace{-2ex}

\section{Conclusions}

We have characterized distinctions between the applicative and
declarative paradigms in functional programming.
We have noted that relational programming exists only in declarative form.
We have developed a relational algebra (TRA) useful as a mathematical 
basis for applicative relational programming.
The operations of TRA are chosen
in such a way that the definite clauses of Prolog have a
simple translation. 

\paragraph{Relations can be first-class citizens.}
One of the most powerful paradigms in computing is ``functions as
first-class citizens'' \cite{sto77}. This in reaction to languages where
functions are more restricted in their use than, say, numbers. 
Certain functional programming
languages have demonstrated the advantages of having functions as
first-class citizens.
Prolog, the only relational programming language, does not
have relation-valued expressions of any kind.
We have shown that relations can be first-class citizens.

\paragraph{Modular logic programming.} We have shown how to support modularity
and local scoping for relations, which is a deficiency in declarative
logic programming. We have devised a mechanism that
facilitates the exportation and importation of
relations between programs.

\paragraph{Higher-order logic is not needed,} at least not to provide
the desirable programming features implied by having functions and
relations as first-class citizens.
In functional programming,
a function is said to be higher-order
when it takes a function as argument or produces one as result. A logic
is said to be higher-order when one can quantify over function or
predicate symbols. These two senses of ``higher-order'' do not coincide.
This is proved by the existence of formalizations of $\lambda $-calculus
in first-order logic.
A constant of logic can denote any
individual, including a function. Therefore, a first-order variable, one
ranging over individuals, can range over functions. In that sense, our
expressions contain first-order variables that range over relations.
This is an important point, as first-order logic
is more tractable, theoretically and practically, than higher-order
logic.

\section{Acknowledgments}

Thanks to J.H.M. Lee for his careful reading of an earlier version of this work.
Generous support was provided by the British
Columbia Advanced Systems Institute, the Institute of Robotics and
Intelligent Systems, the Canadian Institute for Advanced Research,
the Laboratory for Automation, Communication and Information Systems
Research, and the Natural Science and Engineering Research Council of Canada.


\end{document}